\documentstyle[11pt,a4]{article}
\newcommand{\ncm}{\newcommand}
\ncm{\be}{\begin{equation}}
\ncm{\ee}{\end{equation}}
\ncm{\bea}{\begin{eqnarray}}
\ncm{\eea}{\end{eqnarray}}
\ncm{\bc}{\begin{center}}
\ncm{\ec}{\end{center}}
\ncm{\om}{\omega}
\ncm{\Dl}{\Delta}
\ncm{\dl}{\delta}
\ncm{\bt}{\beta}
\ncm{\htm}{\hat\mu}
\ncm{\hlf}{\frac{1}{2}}
\begin{document}
\vspace*{-25mm}
\begin{flushright}
HLRZ 95-73\break
WUB 95-44
\end{flushright}
\begin{center}
  { \bf \Large First Lattice Study of  Ghost Propagators\\[2mm] in SU(2)
    and SU(3) Gauge Theories}\footnote[2]{We ackowledge A.~Hulsebos who
      contributed substantially during the early stages of this work.

} \\ 
    \vspace*{10mm} 
     {\large H.~Suman\footnote[3]{On leave from the Energy Research Group,
              University of Damascus, Syria.}}\\[2mm] 
     {\small\it Fachbereich Physik, University of  Wuppertal, \\
                   Gau\ss{}str. 20, 42097 Wuppertal, Germany}\\
     \vspace*{8mm} 
     {\large K.~Schilling}\\[2mm] 
     {\small\it HLRZ c/o KFA J\"ulich, 52425 J\"ulich, Germany\\ 
      and DESY, Hamburg, Germany}\\ 
    \normalsize \vspace*{20mm} 
     {\bf Abstract}\\ 
    \vspace*{10mm} \parbox{10cm} {We present a numerical study of the
      ghost propagators in Landau gauge for SU(2) and SU(3) gauge
      theories at $\beta$=2.7 and $\beta$=6.0, respectively. Analyzing
      different lattice sizes up to $32^4$, we find small finite
      size effects. Down to the smallest available  momenta, we observe
      no evidence for dipole behaviour of the ghost propagators.\hfill }
\vspace*{6mm}
\end{center}
\section{Introduction}
The gluon propagator, $P$, albeit a non-gauge-invariant quantity, is
considered of prime interest in the quest to gain intuitive insight into
the physics of confinenent in non-Abelian gauge theories.  From a
linearly rising potential, one would argue that $P$ is dominated by a
$1/p^4$ singularity\cite{baker}. This infrared behaviour can be
drastically altered, however, in a scenario of dynamical gluon mass
generation as pointed out in Ref.\cite{cornwall}.

The issue is closely related to the singularity structure of the Green
function of the Faddeev-Popov ghost.  In his seminal paper on gauge
fixing ambiguities {\sc Gribov} \cite{gribov} has dealt with the
implications of the proper choice of integration range (in the
functional gauge field integrals) onto the singularity structure of the
ghost propagator. This is elaborated in detail in a recent comprehensive
paper by {\sc Zwanziger} \cite{zwanziger2} where the theoretical
arguments are presented in quite some detail that lead up to the
prediction of a dipole-type infrared singularity in the Faddeev-Popov
propagator.

It is well accepted by now that the lattice laboratory provides a
valuable tool to the heuristic study of this highly intricate
non-perturbative scenario. Indeed, lattice investigations of the gluon
propagator in Landau gauge did reveal a momentum dependent effective
mass for the gluon\cite{mandula,bernard}. It should be remembered,
though, that -- in view of the very Gribov ambiguities -- such lattice
results rely deeply on the quality and efficiency of gauge fixing
procedures on the lattice \cite{bernard,all}.

In this letter, we wish to complement the previous gluon propagator
investigations by presenting a first numerical study of the ghost
propagator via direct lattice simulation.

\section{Faddeev-Popov Operator}
The ghost fields can be defined on the lattice\footnote{A very useful
collection of the relevant lattice formula related to the ghost fields
can be found in Ref. \cite{zwanziger2}.} in a similar manner as in the
continuum. The ghost propagator $G$ is given in terms of the inverse of
the Faddeev-Popov operator $M$:
\be M = -\nabla D(U), \ee
where $D(U)$ stands for the covariant derivative on the lattice.  It has
a diagonal form in the algebra space, being an average over field
configurations:
\be G(x-y)\dl^{ab} \equiv \Bigl<M^{-1\,ab}(x,y) \Bigr>.
\label{Gprop} \ee
In a Landau gauge fixed configuration the action of the operator $M$ on
an arbitrary element $w$ in the algebra space ${\cal A}$ of the gauge
group SU(N) is given by
\bea (M\om)^a(x)& = & \sum_\mu\Bigl\{ S^{ab}_\mu(x)\,[\om^b(x) -
\om^b(x+\htm)] \;-\;(x\leftrightarrow x-\htm) \nonumber \\ & & - \; \hlf
f^{abc}\,[A^b_\mu(x) \om^c(x+\htm) -
A^b_\mu(x-\htm)\om^c(x-\htm)\,]\;\Bigr\}.  \eea
Here $S$ is a linear functional
\be S^{ab}_\mu(x) = -\hlf \mbox{Tr}\,\Bigl[ (\tau^a\tau^b+\tau^b\tau^a)
\, (U_\mu(x) + U^*_\mu(x))\,\Bigr]. \ee
$A^a_\mu$ is the gluon field and the Pauli matrices $\tau$ are used to
span an antihermitian basis of the linear space ${\cal A}$.

\section{The Simulation}
\paragraph{Run Parameters.}
We ran simulations on a series of lattices, whose parameters are
summarized in Table \ref{lat}.\\[4pt]

\begin{table}[hbt] \begin{center} \begin{tabular}{cccc}
\hline & size & $\beta$ & \# Configurations \\ \hline\hline & $16^4$ &
       2.7 & 50 \\ \raisebox{1.5ex}[-1.5ex]{SU(2)} & $32^4$ & 2.7 & 15
       \\ \hline & $8^4$ & 6.0 & 64 \\ SU(3) & $16^3\times 32$ & 6.0 &
       30 \\ & $24^4$ & 6.0 & 10 \\ \hline
\end{tabular} \end{center} 
\caption{\label{lat}\em Lattices used in this simulation.}
\end{table}

The SU(2) gauge configurations were generated using the
Kennedy-Pendleton heatbath algorithm. 2000 updates were discarded for
thermalization, while gauge fixing and ghost propagator calculations
took place every 100 updates. For SU(3), we applied a hybrid algorithm
of Cabibbo-Marinari heatbath and Creutz overrelaxation steps with mixing
probability 1:5. The configurations have been analyzed every 250 sweeps
after 2000 thermalization steps.

\paragraph{Gauge fixing.} 
Gauge fixing was done by minimizing the functional
\be F_U[g] = \sum_x \sum_\mu \left( 1 - \frac{1}{N} U^g_\mu(x) \right);
\qquad U^g_\mu(x) = g(x)\, U_\mu(x)\, g^{\dagger} (x+\hat{\mu}). \ee
This lattice condition is slightly stronger than the conventional Landau
condition in continuum theory:
\be \partial_\mu A^\mu = 0.\label{lan}\ee
The minimization procedure may be carried out by use of one of the
standard relaxation algorithms which will drive the system to one of its
local minima of $F_U[g]$, thus delivering one of the possible Gribov
copies.

We aim to achieve configurations lying in the fundamental modular region
$\Lambda$, which is given by the set of absolute minima of $F_U[g]$ on
all gauge orbits. The way to do that is to modify the gauging algorithm
and to mobilize the system such as to allow -- in a gentle manner -- for
escape from the attraction of its current closest minimum. To that end
we applied two approaches: (a) the simulated annealing (SA) for SU(2)
and (b) the stochastic overrelaxation (SOR) algorithms for SU(3):

(a) The idea of the annealing algorithm is most easily exposed in the
language of spin models: Minimizing $F_U[g]$ may be viewed as retrieving
the ground state of a spin system, with action being given by
\be S_U(g) = -\beta_s F_U[g]. \ee
In this picture, the ground state may be reached by performing Monte
Carlo sweeps on the `spin' degrees of freedom $g$.

Let us give our prescripton of the annealing algorithm in detail: (1)
start at a spin model coupling $\bt_s = 0.20$, and perform 10 Creutz or
Kennedy-Pendleton heatbath updates on the spin model, at given $\bt_s$;
(2) increase $\bt_s$ in steps of $\Dl\bt_s = 0.20$ and keep updating
until $\bt_s = 25.0$ is reached; (3) at this point turn to ordinary
relaxation.

This gauging procedure was found to be very efficient on both $16^4$ and
$32^4$ lattices.

(b) In the stochastic overrelaxation, a ``wrong'' gauge transformation
is applied once in a while, i.e. with probability $w$, during the
iteration process. If $w$ is sufficiently high, the system may be driven
away from a local minimum.  We applied this method in conjunction with
the Los Alamos algorithm \cite{gupta} after mapping the lattice onto a
checker board basis. We have set the probability $w$=0.7. Our criterion
for achieving the Landau condition was
\be \left| \partial_\mu A^\mu \right|^2 \, < \, 10^{-6}. \ee
On our largest SU(3) lattice, $24^4$, about 1200-1400 iterations were
needed.

\paragraph{ Inversion of $M$.} In  Landau gauge, $M=-\nabla D =
-D\nabla$ is a singular matrix. It annihilates all zero modes on the
left and on the right hand sides. Therefore, in order to compute its
inverse, we have to separate the zero modes and perform the inversion on
its regular part only. $M$ is a real symmetric matrix acting on the
algebra space ${\cal A}$ of the gauge group, which may be decomposed
into two subspaces:
\be {\cal A} = {\cal A}_0 \oplus {\cal A}_1 \ee
such that
\be M\,{\cal A}_0 = 0 \qquad\mbox{and}\qquad M\,{\cal A}_1 \ne 0\quad
(=0 \mbox{ only for the null vector)}.\ee
We start with the simple observation that the vector $Mv$ lies in ${\cal
  A}_1$ for arbitrary $v$ and compute $G$ by solving the linear equation
\be M\, M\, v = M\, s, \label{inversion}\ee
using the conjugate gradient (CG) and minimal residuum (MR) as standard
iterative algorithms. Both methods start off from some initial guess
vector $v_0$ and achieve an improved approximation to the solution by
adding a vector $dv_i$ to the current value $v_i$ at each iteration
step: $v_i \rightarrow v_{i+1} = v_i + dv_i$. Since Eq. \ref{inversion}
ensures that both the starting vector and the source belong to ${\cal
A}_1$, $v_i$ and $dv_i$ will lie also in ${\cal A}_1$. The iteration
will therefore converge to the unique solution within ${\cal A}_1$.

The solution we obtain in this manner coincides with the ghost
propagator restricted to the subspace ${\cal A}_1$ orthogonal to the
zero modes.  Hence, we can study its behaviour in momentum space, down
to (and including) the smallest non vanishing momentum,
$p_{min}=2\pi/L$.

\section{Results}
We calculate the ghost propagators on local sources at $x_0=({L\over
2},{L\over 2},{L\over 2},0)$ and choose the arbitrary normalization
$G(x_0)$=1 on the lattices quoted in Table \ref{lat}.  Then we perform
the Fourier transforms, for different values of lattice momenta:
\be G(p) = \sum_x G(x)\, e^{-ip_\mu (x_\mu - x^0_\mu)}, \qquad p_\mu =
    \frac{2\pi}{L}k_\mu, \quad k_\mu=0,..,L-1.  \ee
In the practical implementation, we keep two components at $k_\mu=0$,
and vary the remaining two components, covering their entire available
kinematical range.

Figure \ref{su3} shows the resulting propagators $G(p)$ in the case of
SU(3), for the lattices $8^4$, $16^4$, $16^3\times 32$, and $32^4$.  The
abscissa is given in terms of the appropriate lattice momentum $q$
\be q(p) = 2\,\sqrt{\sum_\mu \sin^2(\frac{p_\mu}{2})}. \ee

Some observations can be made from the ghost propagators in momentum
space:
\begin{enumerate}
\item On a given lattice the fluctuations of the data points appear to
be small. In the presence of Gribov copies, however, one would expect to
encounter large fluctuations due to gauge dependence. We therefore
conclude that we should not worry about Gribov copies.
\item On the different lattices all data points above
$p_{min}\!=\!2\pi/L$ collapse nicely to a universal curve. Notice in
particular the near-coincidence of the two points at $p\!  =\!2\pi/16$
which are measured in the short and long direction of the $16^3\times
32$ lattice.\hfill\break Looking at the smallest lattice $8^4$, we find
the data point $G(p\!=\! 2\pi /8)$ to lie above the corresponding point
from the $24^4$ lattice by several standard deviations. This tells us
that finite size effects shift $G$ upwards rather than downwards.

\item
Down to the smallest momentum on the $24^4$ lattice, $p\!=\! 2\pi/24$,
the scaling curve increases as momentum decreases. From the
non-symmetric lattice, $16^3\times 32$, we can extract one data point
corresponding to $p_{min}\!=\! 2\pi/32$. This point is found to lie
below its neighbours. It is unlikely that this is due to finite size
effects as the latter would enhance $G(p)$.

Therefore we are tempted to interpret this point as a genuine signal,
that $G(p)$ has a maximum somewhere inside the interval $ [{2\pi\over
32}, {2\pi\over 24} ]$.
\end{enumerate} 
It is comforting to observe a similar structure in our SU(2) data, at
$\beta\!=\! 2.7$ (see Fig.\ref{su2}). The $16^4$ and $32^4$ data
exhibits, after due normalization in configuration space, small finite
size effects.  Again, we observe (a) agreement of $G(p\!=\! 2\pi/16)$ as
measured on the $16^4$ and $32^4$ lattices and (b) a turnover behaviour
of $G(p)$ at very small values of $p$.\hfill\break

We are thus led to the conclusion that there is no direct evidence for
dipole dominance in our data.

%
%

One may expand $G(p)$ in powers of $1/q(p)^2$
\be G(p) = {a\over q^2} + {b\over q^4} + \dots, \label{fitfcn}\ee
and fit the data to the lowest terms. In the region of small momenta, we
were forced to include terms up to ${\cal O}(q^{-8})$, in order to match
the data. However, such fitting implies large cancellations of terms
inducing substantial instabilities.

In the large momentum regime , $q\ge \sqrt{2}$, the data can be
described by the first two terms in Eq.\ref{fitfcn}. The fit yields the
parameters quoted in Table \ref{fit_data}.  Obviously, the quadratic
term $\sim 1/q^2$ accounts for most of the tail in the distribution
$G(p)$, the effective $\simeq 1/q^4$ contribution being determined
within 10\% accuracy.\\[4pt]

\begin{table}[hbt] \begin{center} \begin{tabular}{ccccccc}
\hline & & a & b \\ \hline\hline & $8^4$ & 0.746(8) & 0.276(29) \\ SU(3)
       & $16^3\times 32$(a) & 0.749(9) & 0.227(33) \\ & $24^4$ &
       0.744(10) & 0.256(40) \\ \hline\hline SU(2) & $16^4$ & 2.073(12)
       & 0.267(35) \\ & $32^4$ & 2.084(17) & 0.299(45) \\ \hline
\end{tabular}
\caption{\label{fit_data}\em Fit parameter for the
$q\ge\protect\sqrt{2}$ range, according to the ansatz
\protect\ref{fitfcn}.}
\end{center}\end{table}

\section{Conclusions}
We have studied, for the first time, the ghost propagator for SU(2) and
SU(3) pure gauge theories.

It appears that the gauge fixing algorithms used in the present study
suffice to suppress the effects of Gribov copies. The performance of the
SA algorithm for SU(2) is comparable to the ordinary relaxation
algorithm in CPU time, but its has the advantage of finding the absolute
minimum in a single run. The SOR, in combination with the Los Alamos
algorithm, reduces the number of required iterations by nearly a factor
4, depending on $w$ and on the lattice size. The extra cost for applying
the ``wrong'' gauge transformation is negligible in comparison with the
profit in iteration numbers.

The ghost propagators appear to be only little affected by finite size
effects as we can verify from comparing the data from lattices of
different extensions, at a given $\beta$.

In the infrared regime -- measuring down to momentum values of 0.39 GeV
and 0.86 GeV in SU(3) and SU(2), respectively -- we do not confirm the
expected $1/q^4$ behaviour. Near the boundary of the first Brillouin
zone, $p\rightarrow \pi/L$, $G(p)$ is dominated by $1/q^2$.

\vspace*{4mm} {\bf Acknowledgments}\\[2mm]
We thank D. Zwanziger, who triggered this research, for useful
discussions. This work was supported partially by DFG grants Schi
257/3-3, Schi 257/5-1, Pe 257/1-4 and EU network CHRX-CT92-0051 and by
the Energy Research Group, University of Damascus, Syria.

\vspace*{6mm} {\Large\bf Figure Captions}\\
\vspace*{-10mm}
\begin{figure}[h]\begin{center} 
\caption{\protect\label{su3} $G(q)$ for SU(3) at $\beta\!=\! 6.0$ on
different lattice sizes. The curve represents the fit to the $24^4$-data
according to the ansatz (Eq.\protect\ref{fitfcn}) in the range
\hbox{$q\!\geq\!\protect\sqrt{2}$}. Note that the full $q$-range in our
simulation is given by $0\!\leq\! q\!\leq\! 2\protect\sqrt{2}$.  }
\end{center}\end{figure}

\vspace*{-15mm}
\begin{figure}[h] \begin{center}
\caption{\protect\label{su2} The same as Fig. \protect\ref{su3}, but for
SU(2) at $\beta$=2.7. The fit shown is for the $32^4$ lattice.}
\end{center} \end{figure}

\clearpage\newpage\thispagestyle{empty}\phantom{hallo}
\includegraphics{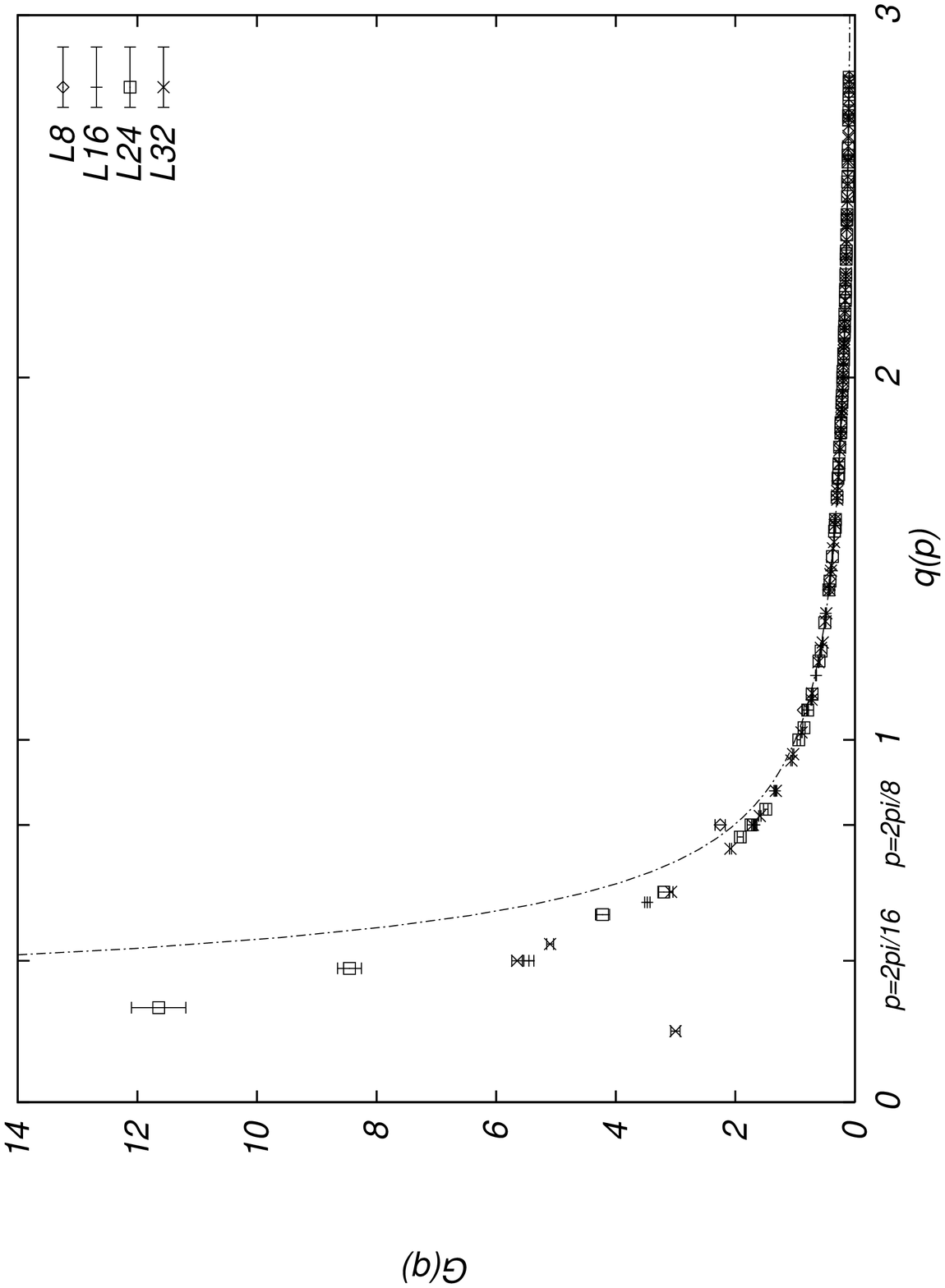}

\clearpage\newpage\thispagestyle{empty}\phantom{hallo}
\includegraphics{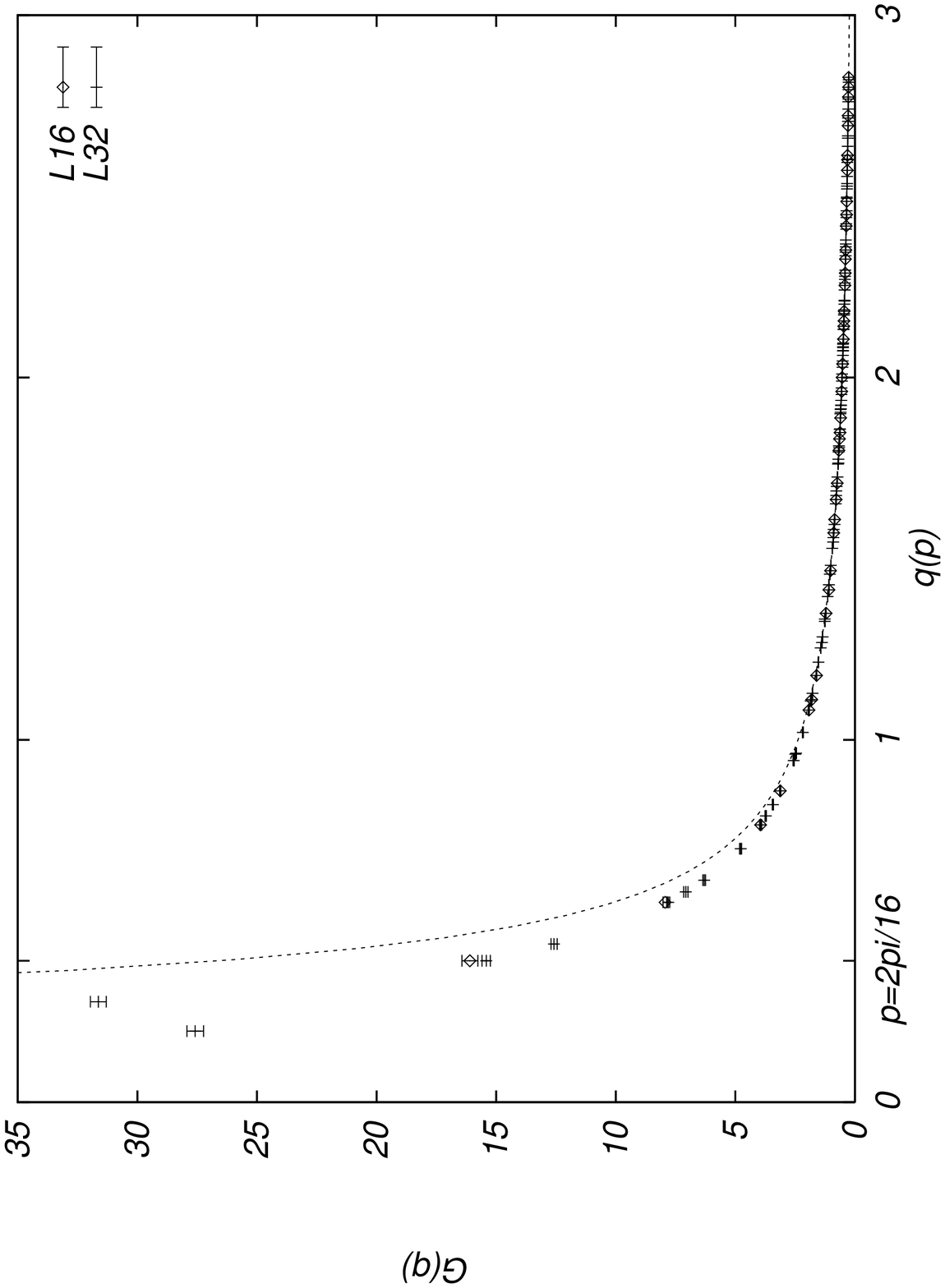}

\end{document}